------------------------------------------------------------------------------
\documentstyle[12pt,a4]{article}

\newcommand{\be}{\begin{equation}}
\newcommand{\ee}{\end{equation}}
\newcommand{\bea}{\begin{eqnarray}}
\newcommand{\eea}{\end{eqnarray}}
\newcommand{\unop}{1\!{\rm I}}

\title{On the relations between osp(2,2) and the \\
quasi exactly solvable systems}

\author{Y. Brihaye \\
Department of Mathematical Physics\\
University of Mons\\
Av. Maistriau, B-7000 MONS, Belgium.\\
Stefan Giller{$^{*}$},
       Piotr Kosinski \thanks{$^{\dagger}$ Work supported by grant $n^o$
KBN 2P03B07610}\\
Department of Theoretical Physics\\
University of Lodz\\
Pomorska 149/153, 90-236 Lodz, Poland}
\begin{document}

\begin{titlepage}
\maketitle
\thispagestyle{empty}
\begin{abstract}
By taking a product of two sl(2) representations,
we obtain the differential operators preserving
some space of polynomials in two variables. This 
allows us to construct the representations of osp(2,2)
in terms of matrix differential operators in two variables. 
The  corresponding operators provide the building blocks 
for the construction of quasi exactly solvable systems
of two and four equations in two variables. Some generalisations
are also sketched. The peculiar labelling used for the generators
allows us to elaborate a nice deformation of osp(2,2). This gives an
appropriate basis for analyzing the quasi exactly solvable systems
of finite difference equations.     
\end{abstract}
\end{titlepage}

\hfill {\it In memory of our friend S. Malinowski}
\vskip 2. cm 
\section{Introduction}
The number of quantum mechanical problems for which the spectral equation can
be solved algebraically is rather limited. It is therefore not surprising
that the quasi exactly solvable (QES) equations 
\cite{tur1}, \cite{ush} attract some attention.
For these equations, indeed, a finite number of eigenvectors can be obtained
by solving an algebraic equation. The study of QES equations has motivated
the classification of finite dimensional real Lie algebras of first order
differential operators. The algebras which, in a suitable representation,
preserve a finite dimensional module of smooth real functions
are particularly relevant for QES equations. The case of
one variable was addressed and solved some years ago \cite{tur2}. 
There is, up to
an equivalence only one algebra, for instance sl(2), acting on the space of
polynomials of degree at most $n$ in the variable.
For two variables, the classification is more involved. It is described
respectively in \cite{gko1} and \cite{gko2} for complex and real variables.
The corresponding real QES operators finally 
emerge in seven classes summarized
in table 7 of \cite{gko2}.

The natural next step is to classify the QES systems of two 
equations \cite{turshif,bk}. It is known that, in general, the 
underlying algebraic structure is not a (super) Lie algebra;
in those few cases where the algebra is a Lie algebra, 
it is generically sl(2)$\times$sl(2) or osp(2,2) \cite{bk}.
The first case is rather trivial while the second one seems
to be worth to be studied in more details.
This was discussed in \cite{turshif,bk}; the present paper
is devoted to a further study of the relations of osp(2,2)
with QES systems.

In the second section, we present some new aspects of 
the representations of the Lie algebra sl(2)
in terms of differential operators. As a byproduct we obtain a simpler
description of some of the QES operators 
classified in \cite{gko2}. Thereafter we construct the realizations
of osp(2,2) in terms of partial differential operators of two real 
variables.
This provides the algebraic basis for the description of certain QES systems
of two or four equations in two variables.
The labelling we use for the generators of osp(2,2)
simplifies considerably the structure of the algebra.
Taking advantage of this labelling,
we are able to find a deformation of osp(2,2) similar, in the
spirit, to the Witten-Woronowic deformation of sl(2) \cite{witt,woro,zachos}.
For this deformed algebra, we also find the representations  
that we express in terms of finite difference operators.
The normal ordering rules obeyed  by these operators
are more appropriate for the classification of finite difference
QES operators than the ones given in a previous paper \cite{bgk}.

\section{Representations of sl(2)}

We first discuss the realizations of sl(2) which will
be usefull in the next sections.
In the case of one variable, the basic  QES operators read
\cite{tur1},\cite{tur2}
\be
j_- (x,m) = {d\over {dx}}\quad , \quad
j_0 (x,m)  = x{d\over {dx}} - {m\over 2}\quad , \quad 
j_+ (x,m) = x^2 {d\over{dx}} - mx
\label{sl2}
\ee
When $m$ is an integer these operators
preserve the vector space, say $P(m)$,
of polynomials of degree at most $m$
in the variable $x$.
They obey the commutation relations of the Lie algebra sl(2).

From (\ref{sl2}), we can define the operators corresponding
to the product of two representations. That is to say, 
if $x$ and $y$ denote two independent variables  
\be
\label{sl2d}
j_{\epsilon}(x,m;y,n) = j_{\epsilon}(x,m) + j_{\epsilon}(y,n)\quad ,
\quad \epsilon = \pm, 0
\ee
Clearly, the operators (\ref{sl2d})
obey the same commutation relations as (\ref{sl2}).
They act on the vector space, say $P(m,n)$ of polynomials of degree at most
$m$ (resp. $n$) in the variable $x$ (resp. $y$).
However, their action  on $P(m,n)$ is not irreducible. 
One can show easily that the operators (\ref{sl2d})
preserve irreducibly the subspace of $P(m,n)$ defined by
\be
M(m,n) = {\rm Span} \left\lbrace 
({\partial\over{\partial x}} + {\partial\over {\partial y}})^k
x^m y^n \quad , \quad 0 \leq k \leq m+n\right\rbrace
\label{sousespace}
\ee
This vector space is the eigenspace of the Casimir operator
corresponding to the representation of highest spin in the 
decomposition of $P(m,n)$ in subspaces irreducible with respect to
the action  of (\ref{sl2d}).
The other irreducible representations
result as similar structures with different values of $m$ and $n$;
it is therefore sufficient to deal with (\ref{sousespace}).

Alternatively, $M(m,n)$  can be seen as the kernel  of the operator
\be
K(x,y,m,n) = (x-y) {\partial\over{\partial x}} {\partial \over{\partial y}}
+ n{\partial \over{\partial x}} - m {\partial \over{\partial y}}
\ee
acting on $P(m,n)$.
Remark that $M(m,0)$  (resp. $M(0,n)$) is isomorphic to $P(m)$ (resp. $P(n)$);
it that case, the one dimensional operators (\ref{sl2})
are recovered from (\ref{sl2d})  by ignoring
the partial derivative $\partial \over \partial y$ (resp.  
$\partial \over \partial x$).

\subsection{QES operators in two variables}

The QES operators in two real variables are classified in Ref. \cite{gko2};
the authors summarize the seven possible hidden algebras 
in their table 7.
The operators labelled (1.4), (1.10) and (2.3) in this table
are studied independently in Refs. \cite{turshif}, \cite{tur3}. 
The operators labelled (1.1) in the table appear to be new; 
in particular they lead to the only case
for which the invariant module is not 
manifestly a space of polynomials in the two variables.  
In the following we show that the 
formulation (1.1) can be simplified and related
to the algebra (\ref{sl2d}) by means of a suitable change of function.

The operators (1.1) in table 7 of ref.\cite{gko2} read
\begin{eqnarray}
\tilde j_- &=& {\partial\over{\partial x}} + {\partial \over{\partial y}},
\nonumber \\
\tilde j_0 &=& x{\partial\over{\partial x}} + y {\partial\over{\partial y}},
\nonumber \\
\tilde j_+ &=& x^2 {\partial\over{\partial x}} + y^2{\partial\over{\partial y}}
+ {n\over 2} (x-y)
\label{olver11}
\end{eqnarray}
They preserve the space $\tilde M(m,n)$ defined as
\be
\tilde M(m,n) = {\rm Span} \left\lbrace (x-y)^{m+{n\over 2}-k} R_k^{m,n}
\left({x+y\over {x-y}}\right)\quad , \quad 0 \leq k \leq 2m+n\right\rbrace
\ee
\be
R_k^{m,n}(t) = {d^k\over{dt^k}} (t-1)^{m+n}(t+1)^m
\ee
Our observation is summarized by the following two formulas
\be
(x-y)^{m+{n\over 2}} \  \tilde j_{\epsilon} \ (x-y)^{-m-{n\over 2}} = 
j_{\epsilon}(x,m) + j_{\epsilon}(y,m+n)
\ee
\be
(x-y)^{m+{n\over 2}} \tilde M(m,n) = M(m,m+n)
\ee
In other words the algebra (\ref{olver11}) is equivalent to the
algebra (\ref{sl2d}) (up to a suitable redefinition of $n$
into $m+n$).
\par The advantage of the new formulation is twofold. 
First the relevant operators form an sl(2) 
diagonal subalgebra of the sl(2)$\otimes$ sl(2) algebra generated by 
\be
j_{\epsilon}(x,m)\quad , \quad j_{\epsilon}(y,n)\quad , \quad \epsilon =
\pm, 0
\ee
In this respect the form (\ref{sl2d}) of the operators (\ref{olver11})
is clearly related to the fundamental operators (\ref{sl2})
and  is easy to generalize to the case of $V$ variables.  
Second, the vector space of the representation, 
i.e. $M(m,n)$, is a space of polynomials like in
all the other cases of the classification of Ref.\cite{gko2}.
In the next sections, we discuss some possible extensions 
of the algebra (\ref{sl2d}) into graded algebras. The
corresponding operators  are related to  systems of QES  equations. 

\subsection{Imbeddings into gl(2)}

In the next sections, we will put the emphasis on the  
classification of the operators which preserve the direct sum
\be
M(m,n)\oplus M(m+\Delta,n+\Delta') \ \ 
\label{sumgen}
\ee
where $m,n$, $m+\Delta, n+ \Delta'$
are positive integers.
The operators (\ref{sl2d}) are crucial for this task; 
however we find it convenient to label them as follows. 
For $\mu,\nu \in \{0,1 \}$ we define 
\be
J_{\mu}^{\nu} (x,m,m+\Delta;y,n,n+\Delta') = 
{\rm diag} (j_{\mu}^{\nu}(x,m;y,n), j_{\mu}^{\nu} (x,m+\Delta;y,n+\Delta'))
     - {T \over 2} \delta_{\mu}^{\nu} 
%
%
\label{gl2}
\ee
where $T$ is a constant $2 \times 2$ diagonal matrix while
\bea
&j_0^0(x,m;y,n) &= j_0(x,m;y,n) \nonumber \\
&j_1^1(x,m;y,n) &= -j_0(x,m;y,n) \nonumber \\
&j_0^1(x,m;y,n) &= j_-(x,m;y,n)   \nonumber \\
&j_1^0(x,m;y,n) &= - j_+(x,m;y,n)  
\eea
The operators (\ref{gl2})
obey the commutation relations of the Lie algebra gl(2) :
\be
[ J_{\mu}^{\nu} , J_{\alpha}^{\beta}] = 
  \delta_{\mu}^{\beta} J_{\alpha}^{\nu} 
- \delta_{\alpha}^{\nu} J_{\mu}^{\beta} \ \ ,
\label{gl2com}
\ee
(for shortness we drop the dependence on $x,m,\ldots$)
defining a (reducible) representation of this Lie algebra
on the vector space  (\ref{sumgen}). 

\section{Representations of osp(2,2)}

We first consider the $2\times 2$ matrix 
operators preserving the direct sum
\be
M(m,n) \oplus M(m+1,n) \ \ .
\label{sum1}
\ee
In addition to the diagonal generators $J_{\mu}^{\nu}(x,m,m+1;y,n,n)$,
defined above, we have to construct  the off diagonal ones. 
They can be formulated in terms 
of the matrices $\sigma_{\pm} \equiv (\sigma_1 \pm i \sigma_2)/2$
($\sigma_a$, a=1,2,3 are the Pauli matrices) and of the following
differential operators 
\begin{eqnarray}
q_0(x,m;y,n) &=&
  {1\over{m+1}} (m+n+1+(x-y){\partial\over{\partial y}}) \nonumber \\
q_1(x,m;y,n) &=& 
-{1\over{m+1}} \left((m+1)  x+ny+y(x-y){\partial
\over{\partial y}}\right)  \label{q} \\
\overline q_1(x,m) &=& {\partial\over{\partial x}} \nonumber \\
\overline q_0(x,m) &=& (x {\partial\over{\partial x}} - m - 1)
 \label{qb}
\end{eqnarray}
The coming proposition allows to classify the operators under consideration.

\vspace{0.3cm}
\noindent {\bf{Proposition}}

\vspace{0.3cm}
\noindent
{\it The linear operators preserving the vector space (\ref{sum1}) 
are the elements of
the enveloping algebra generated by the  
eight operators}
\be
J_{\mu}^{\nu}(m,n,1,0) \ \  , \ \ \mu, \nu = 0,1
\label{op1}
\ee
\be
Q_{\mu} = q_{\mu}(x,m;y,n) \sigma_- \qquad  , \qquad {\mu}=0,1
\label{op2}
\ee
\be
\overline Q^{\mu} = \overline q^{\mu}(x,m) \sigma_+ \qquad ,
\qquad {\mu}=0,1
\label{op3}
\ee

The eight operators (\ref{op1}),(\ref{op2}),(\ref{op3}) 
acting on the space (\ref{sum1}) lead to an irreducible
representation (with dimension = $2m+2n+3$)
of the graded Lie algebra osp(2,2).
The labelling of the generators,  together with the matrix $T$ in 
(\ref{gl2}) of the form
\be
  T = {\rm diag} (m+n+\Delta+\Delta'+1 , m+n+1)
\label{top}
\ee
results in a  
particularly concise form of the structure constants. 
In addition to (\ref{gl2com}), we find
\bea
&[ J_{\mu}^{\nu} , Q_{\alpha} ] &=  
 \delta_{\mu}^{\nu} Q_{\alpha} - \delta_{\alpha}^{\nu}  Q_{\mu} \\
&[ J_{\mu}^{\nu} , \overline Q^{\alpha} ] &=  
  \delta^{\alpha}_{\mu}  \overline Q^{\nu} -
 \delta_{\mu}^{\nu} \overline Q^{\alpha}  \\
&\{ Q_{\mu} , \overline Q^{\nu} \} &=  J_{\mu}^{\nu} \\
&\{ Q_{\mu} ,  Q_{\nu} \} &=
 \{ \overline Q^{\mu} , \overline Q^{\nu} \} = 0 
\eea
In particular $Q_{\mu}$ (and $\overline Q^{\mu}$)
transforms as a doublet under the adjoint action of the gl(2)
subalgebra generated by the four $J$'s.

The relationship  between the super Lie algebra  osp(2,2)
and the differential operators (of one variable) 
preserving P(m)$\oplus$ P(m+1)
was first noticed in  \cite{turshif}.
The differential operators used in \cite{turshif}
can be recovered from ours by setting $n=0$ and dropping
all derivatives ${\partial \over \partial y}$. 
Our labelling of the generators is, however, different.

The finite dimensional representations 
of osp(2,2) \cite{snr} can be expressed in
terms of differential operators of one variable \cite{bgk}. 
In this kind of approach, the generic,
finite dimensional irreducible representation 
appears as acting on the vector space
\be
P(m) \oplus P(m+1) \oplus P(m-1) \oplus P(m).
\ee
The representations of osp(2,2) 
can be formulated also in terms 
of the two variables differential operators constructed above.
In particular the generic irreducible representation
can be constructed equally well on the vector spaces 
\be
V_I = M_{m,n} \oplus M_{m+1,n}\oplus M_{m-1,n}\oplus M_{m,n}
\ee
\be
V_{II} = M_{m,n} \oplus M_{m+1,n}\oplus M_{m,n+1}\oplus M_{m,n}
\ee
The associated representations are equivalent 
(in aggrement with \cite{snr}) but 
the expressions of the generators 
in terms of the partial derivatives is quite different. 
The proof of the equivalence between the representations acting on $V_I$ and 
on $V_{II}$ relies on the fact that the operators (\ref{gl2}) are invariant
under the double substitution 
$m \leftrightarrow n$ and $x \leftrightarrow y$.
Therefore the operators $q_{\mu}(y,n;x,m)$ 
(resp. $\overline q^{\mu}(y,n)$) behave exactly as $q_{\mu}(x,m;y,n)$
(resp.  $\overline q^{\mu}(x,m)$) under the adjoint action of (\ref{gl2}).

\section{More graded algebras}

We now put the emphasis on the operators preserving the vector space
\be
 M_{m,n} \oplus M_{m+\Delta,n + \Delta'} \quad \ \ . \ \ 
\ee
The relevant diagonal operators can be chosen according to (\ref{gl2}).
The construction of the off diagonal ones depends on the relative signs
of $\Delta$ and $\Delta'$.

\subsection{case $\Delta,\Delta' \geq 0$}
The operators connecting the vector space
$M(m,n)$ with $M(m+\Delta,n+\Delta')$ (and vice versa)
can be formulated in terms of products
of  operators (\ref{q}) (and  (\ref{qb})) where the indices 
$m$ and $n$ in the different factors are appropriately shifted .
The following identities allows one to deal with the 
ambiguities of ordering of
the different factors:
\begin{eqnarray}
&q_b(x,m+1;y,n) \ q_a(x,m;y,n) &=q_a(x,m+1;y,n) q_b(x,m;y,n) \nonumber \\
&q_b(y,n;x,m+1) \ q_a(x,m;y,n) &=q_a(y,n;x,m+1) q_b(x,m;y,n) \nonumber \\
&q_b(y,n;x,m+1) \ q_a(x,m;y,n) &= q_a(x,m;y,n+1) q_b(y,n;x,n) 
\label{iden}
\end{eqnarray}
One can use these identities in order to define the operators 
\be
q(x,[a_k];y,[b_l]) \equiv 
\prod_{l=1}^{\Delta '} 
\prod_{k=1}^{\Delta}
q_{b_l}(y,n+l-1;x,m+ \Delta) 
q_{a_k}(x,m+k-1,n)
\label{qprod}
\ee
symmetrically  in the multi indices 
\bea
&[a_k] \equiv (a_1, \ldots , a_{\Delta '}) \ \ \ , \ \ \ &a_i = 0,1
\nonumber \\
&[b_l] \equiv (b_1, \ldots ,  b_{\Delta }) \ \ \ , \ \ \ &b_i = 0,1
\eea
The operators (\ref{qprod}) connect $M(m,n)$ with $M(m+\Delta,n+\Delta')$
and
\be
Q([a_k],[b_l]) =  q(x,[a_k];y,[b_l]) \sigma_- 
\ee
are the counterparts of the  operators (\ref{q}).
Using the remarks made at the end of the previous
section, one can convince that the operators (\ref{qprod})
transform according to the representation of spin $\Delta + \Delta'$
under the adjoint action of the gl(2) represented via (\ref{gl2}).

The operators $\overline Q([a_k],[b_l])$, proportional to
$\sigma_+$,  can be constructed 
in exactly the similar way as for the $Q([a_k],[b_l])$.
Identities like (\ref{iden}) exist among the $ \overline q_a$.
The complete algebra (which is non linear)
 can be obtained following the same lines
as in \cite{bn}.

\subsection{The case M(m,n) $\oplus$ M(m+1,n-1)}

If we consider $\Delta$ and $\Delta'$ of opposite signs,
the operators preserving (\ref{sumgen}) do not represent
a Lie super algebra even for 
$\vert \Delta \vert = \vert \Delta' \vert =1$
We studied the operators preserving the vector space 
$M(m,n) \oplus M(m+1,n-1)$
and observed that the underlying algebric stucture is different from those 
obtained in \cite{bn}. Again,
 $J_{\mu}^{\nu}(x,m,m+1;y,n,n-1)$ can be used
as a starting point. 
We find it convenient to set $T=0$ in (\ref{gl2})
and add separately the  grading operator 
$T\equiv \sigma_3$.
As far as the off diagonal operators are concerned, we choose
\bea
     &R_{\mu}^{\nu} &= 
\overline q_{\mu}(y,n-1,x,m+1) q^{\nu}(x,m,y,n) \sigma_-
\nonumber \\
     &\overline R_{\mu}^{\nu} &= 
\overline q_{\mu}(x,m,y,n) q^{\nu}(y,n-1,x,m+1) \sigma_+
\label{rrbar}
\eea
These tensors are not irreducible with respect to the adjoint action
of the $J$ generators. 
The traces
\be
R_{\mu}^{\mu} = 
{m+n+2 \over m+1} ((y-x) {\partial \over \partial y} -n) \ \ \ , \ \ \ 
\overline R_{\mu}^{\mu} = 
{m+n+2 \over n+1} ((x-y) {\partial \over \partial x} -m)
\ee
are operators which intertwines the equivalent representations
carried by the spaces $M(m,n)$ and $M(m+1,n-1)$;
they commute with the four operators (\ref{gl2}). 

The  order of the factors $q$ and $\overline q$ entering in
$R$ and $\overline R$ can be reversed by using the identity
\bea
q_{\mu}(y,n-1,x,m+1) q^{\nu}(x,m,y,n) &-&
q^{\nu}(x,m,y,n-1)q_{\mu}(y,n-1,x,m) \nonumber \\
&=& {1 \over m+n+2} R_{\mu}^{\mu}
\eea

The generators $J_{\mu}^{\nu},R_{\mu}^{\nu}, \overline J_{\mu}^{\nu}$
obey   the following commutation 
and anticommutation relations 
\bea
&[ J_{\mu}^{\nu} , R_{\alpha}^{\beta}] &= 
  \delta_{\mu}^{\beta} R_{\alpha}^{\nu} 
- \delta_{\alpha}^{\nu} R_{\mu}^{\beta} \ \ , \nonumber \\
&[ J_{\mu}^{\nu} , \overline R_{\alpha}^{\beta}] &= 
  \delta_{\mu}^{\beta} \overline R_{\alpha}^{\nu} 
- \delta_{\alpha}^{\nu} \overline R_{\mu}^{\beta}
\eea
\bea
\{ R_{\mu}^{\nu} ,  \overline R_{\alpha}^{\beta} \}
&=& {1\over 2} \{ J_{\mu}^{\beta} , J_{\alpha}^{\nu} \}  
+ {T\over 2} (   \delta_{\mu}^{\beta}  J_{\alpha}^{\nu} 
                -  \delta_{\alpha}^{\nu} J_{\mu}^{\beta} ) \nonumber \\ 
&-& {1\over 2} (   \delta_{\mu}^{\nu}  J_{\alpha}^{\beta} 
                +  \delta_{\alpha}^{\beta} J_{\mu}^{\nu} ) 
- {1\over 2}  \delta_{\mu}^{\nu}  \delta_{\alpha}^{\beta}   
\label{corrbar}
\eea
These  relations do not define an abstract algebra.
In order to fulfil the associativity conditions
(i.e. the Jacobi identities), the  anticommutators between 
two $R$ (two $\overline R$), which vanish for (\ref{rrbar}) 
have to be implemented in a non trivial way \cite{bk},\cite{bn};
equation (\ref{corrbar}) indicates that the underlying algebra 
is non linear.

\section{Deformation of osp(2,2)}

If we want to describe in abstract terms the
algebraic structure  underlying the QES finite difference equations,
some deformations of the algebras discussed above seem to emerge
in a natural way.
For scalar equations the relevant deformation was pointed out
some time ago \cite {tur2}.  
In a previous paper \cite{bgk} we presented some 
(finite dimensional) representations
of a deformation of osp(2,2) in terms of 
finite difference operators.
However, we have taken advantage of the ``gl(2)-labelling'' 
used here for the generators of osp(2,2) to construct a new 
deformation of this super Lie algebra.

A deformation of osp(2,2) is known \cite{deguchi} 
(for more general graded algebra see \cite{vinet});
it is such that
the (anti) commutators of some generators close,
within the enveloping algebra,
in terms of transcendental functions of the generators.
The deformation of osp(2,2)  that we present here is constructed in 
the same spirit as the
so called ``Witten type II'' deformation  of sl(2) \cite{witt,zachos}.
That is to say that the structure of the algebra relates 
q-commutators of the generators to linear combinations of them.
The q-commutators and q-anti-commutators are defined respectively
as
\be
   [A , B]_q = AB - q BA \ \ \ \ , \ \ \ \ \{A , B \}_q = AB + qBA
\ee
where $q$ is the deformation parameter.

Our deformation of osp(2,2) is expressed as follows,
\be
  \{ Q_{\mu} , Q_{\nu} \}_{q^{\nu-\mu}} = 0 \ \ \  , \ \ \  
 \{ \overline Q^{\mu} , \overline Q^{\nu} \}_{q^{\nu-\mu}} = 0 
\ee
\be
 \{ Q_{\mu} , \overline Q^{\nu} \} = J_{\mu}^{\nu}  
\label{defj}
\ee
\be
[ J_{\mu}^{\nu} , Q_{\alpha} ]_{q^{\alpha-\nu}} =
q^{{\alpha-\nu-1}\over 2}  
 (\delta_{\mu}^{\nu} Q_{\alpha} - \delta_{\alpha}^{\nu}  Q_{\mu}) 
\ee
\be
[ J_{\mu}^{\nu} , \overline Q^{\alpha} ]_{q^{\mu -\alpha}} =  
q^{{\mu-\alpha-1}\over 2}  
 (\delta^{\alpha}_{\nu}  \overline Q^{\nu} -
    \delta_{\mu}^{\nu} \overline Q^{\alpha}  ) 
\ee
\bea
[ J_{\mu}^{\nu} , J_{\alpha}^{\beta}]_{q^s} &=& 
 q^{{s-r-1}\over 2}(\delta_{\mu}^{\beta} J_{\alpha}^{\nu} 
- q^{r} \delta_{\alpha}^{\nu} J_{\mu}^{\beta}) \nonumber \\
&+& {q-1 \over q^2} (Q_0 \overline Q^0 + Q_1 \overline Q^1)
\delta_{\alpha}^{\nu} \delta_{\mu}^{\beta}
(1-\delta_{\mu}^{\nu} \delta_{\alpha}^{\beta})
\label{gl2def}
\eea
with $s\equiv \nu + \alpha - \mu - \beta$ and
$r \equiv (\nu - \beta)(\mu - \alpha)$.

All Jacobi identities are obeyed by these relations.
The last commutator indicates that the bosonic generators do not 
close into a gl(2) subalgebra (similarly to the  
deformation of ref.\cite{deguchi}).
Neglecting all fermionic operators in the above formulas
leads to one deformation of gl(2) found in \cite{fairlie}. 

It is possible to construct two independent expressions,
quadratic in the generators, which q-commute with the generators.
These ``q-Casimir'' operators read 
\bea
      &C_1 = &Q_0 \overline Q^0 + Q_1 \overline Q^1
             + q J_1^0 J_0^1 - J_0^0 J_1^1 - J_0^0 \nonumber \\
      &C_2 = &(q-1)^2 (Q_0 \overline Q^0 + Q_1 \overline Q^1)
             + q(q-1)^2 J_1^0 J_0^1 + (q-1) J_1^1 -q(q-1)J_0^0 -1 
\eea
and obey the following commutation properties (for i=1,2)
\bea
   &[C_i, J_{\mu}^{\nu} ]_{q^{2(\mu - \nu)}} &= 0 \ \ \ ,  \nonumber \\
   &[C_i, Q_{\mu}]_{q^{2\mu - 1}} &= 0 \ \ \ ,  \nonumber \\
   &[C_i, Q^{\mu}]_{q^{1-2\mu }} &= 0 \ \ \ 
\eea
It follows that any function of the ratio $C_1/C_2$ commute with the generators
and constitute a conventional Casimir.

We further constructed the representations of the algebra above
which are relevant for systems of finite difference QES equations.
To describe them we define a finite 
difference operator  $D_q$.  
\be
       D_q f(x) =  { f(x) - f(qx) \over (1-q)x} \ \ \ , \ \ \ 
       D_q x^n = [n]_q x^{n-1} \ \ \ , \ \ \   
       [n]_q \equiv {1 - q^n \over 1-q}  
\ee 

The simplest of these realizations are characterized by a positive
integer $n$ and  act on the vector space
\be
P(n-1) \oplus P(n) \ \ .
\label{pp}
\ee
 Adopting $x$ as the variable,  
the fermionic generators are represented by
\bea
   &Q_0 = q^{-{n\over 2}} \sigma_- \ \ \ , \ \ \  
   &Q_1 = -x \sigma_- \nonumber \\ 
   &\overline Q^0 = q^{-{n\over 2}} (x D_q - [n]_q) \sigma_+ \ \ \ , \ \ \  
   &\overline Q^1 = D_q \sigma_+ 
\label{repp}
\eea
The bosonic operators can be constructed easily from (\ref{defj}) but
we write them for completeness
\be
J_0^0 = q^{-n} ((x D_q - [n]_q) \unop_2 \ \ \ \ , \ \ \ 
J_1^1 = (-1) {\rm diag } ( q x D_q + 1 , x D_q)
\ee
\be
J_0^1 = q^{-{n\over 2}}  D_q \unop_2 \ \ \ \ , \ \ \ 
J_1^0 = (-1)q^{-{n\over 2}}  {\rm diag } 
( qx(x D_q - [n-1]_q) , x(x D_q - [n]_q) )
\ee
The enveloping algebra constructed with the eight generators above
contains all finite difference operators preserving $P(n-1) \oplus P(n)$.

As in the underformed case, there also exist representations
which act on the vector space 
\be
     P(n) \oplus P(n+1) \oplus P(n-1) \oplus P(n)
\label{pppp}
\ee
The fermionic generators are realized as follows
\be
Q_0 =
\left(\begin{array}{cccc}
0     &0     &0   &0\\
1     &0     &0   &0\\
D_q   &0     &0   &0\\
0     &-D_q  &1   &0
\end{array}\right)
\nonumber
\ee
\be
Q_1 =
\left(\begin{array}{cccc}
0     &0     &0   &0\\
x     &0     &0   &0\\
q^{-n}\delta(n)   &0     &0   &0\\
0     &-q^{-n-1}\delta(n+1)  &x   &0
\end{array}\right)
\nonumber
\ee
\be
\overline Q^0 =
\left(\begin{array}{cccc}
0  &\lambda \delta(n+1) &(1-\lambda q^{n+1})x    &0    \\
0  &0                   &0   &(\lambda q^{n+1}-1)x       \\
0  &0                   &0   &q\lambda \delta(n)         \\
0  &0                   &0   &0   
\end{array}\right)
\nonumber
\ee
\be
\overline Q^1 =
\left(\begin{array}{cccc}
0  &\lambda q^{n+1}D_q  &(1-\lambda q^{n+1})      &0   \\
0  &0                   &0   &\lambda q^{n+1} -1       \\
0  &0                   &0   &\lambda q^{n+1} D_q      \\
0  &0                   &0   &0   
\end{array}\right)
\label{repppp}
\ee
where $\lambda$ is an arbitrary complex parameter and, 
for shortness, we used $\delta(n) \equiv xD_q - [n]_q$. 
The invariance of the operators under the similarity
transformation has been exploited to set $Q_0$ in a form
as simple as possible.
In the limit $\lambda \rightarrow q^{-n-1}$ the representation
(\ref{repppp}) becomes reducible and decomposes 
into atypical ones of the form (\ref{repp}).

The deformation  of osp(2,2) presented above leads to  
simple  normal ordering rules for the generators.
In this respect, it is very appropriate 
for the classification of the finite 
difference operators preserving the space (\ref{pp}) or (\ref{pppp})
and of the corresponding QES systems.
The normal ordering rules 
associated with the deformation used in ref.\cite{bgk} 
are not as transparent since they depend non polynomially
of some operators.

\section{Concluding remarks}

The most interesting examples of QES systems are related to
the algebra osp(2,2), e.g. the relativistic Coulomb problem
and the stability of the sphaleron in the abelian Higgs model
\cite{bk}. Therefore, further realizations of this algebra
in terms of differential operators desserve some attention.
Here, we present realizations formulated in terms of 
differential operators in two variables.
The labelling used for the generators clearly exhibits their
tensorial structure under the gl(2) subalgebra and 
provides very naturally the building blocks for the construction
of series of (non linear) graded algebras preserving some vector
spaces of polynomials.

Witten type deformations attracted recently some attention
(see e.g. \cite{chung} for  osp(1,2)).
The deformation of osp(2,2) presented here is of this type
and it admits representations which are directly relevant for
the study of QES finite difference systems.
The existence of a coproduct for this type of deformation
would allow to adapt the construction of section 2 to
finite difference equations. This is a nice application
of the coproduct that we plan to address in a subsequent paper.

\end{document}